\documentclass[11pt,reqno]{article}
\usepackage{amsmath, amsfonts, amscd, eucal}
\usepackage{latexsym}
\usepackage{amssymb}

\newcommand{\be}{\begin{equation}}
      \newcommand{\ee}{\end{equation}}
\newcommand{\ban}{\begin{eqnarray*}}
       \newcommand{\ean}{\end{eqnarray*}}
\newcommand{\ba}{\begin{eqnarray}}
       \newcommand{\ea}{\end{eqnarray}}

\begin{document}
\newcommand{\pt}{\partial}
\newcommand{\lp}{\langle}
\newcommand{\rp}{\rangle}
\newcommand{\ra}{\rightarrow}
\newcommand{\rat}{\mapsto}
\newcommand{\lra}{\longrightarrow}
\newcommand{\ul}{\underline}
 \renewcommand{\o}[2]{\frac{#1}{#2}}
\newcommand{\hf}{\o{1}{2}}

\newcommand{\qed}{\hspace*{\fill}\rule{3mm}{3mm}\quad}
 \newcommand{\Pf}{\noindent {\em Proof.} }
\newcommand{\Ex}{\noindent {\em Example.} }
\newcommand{\Rk}{\noindent {\em Remark.} }
\newcommand{\Def}{\noindent {\em Definition.} }
\newcommand{\heta}{\hat{\eta}}
\newcommand{\reta}{\overline{\eta}}
\newcommand{\ch}{\mbox{\rm ch}}
\newcommand{\sgn}{\mbox{\rm sgn}}
\newcommand{\tr}{\mbox{\rm tr}}
\newcommand{\Tr}{\mbox{\rm Tr}}
\newcommand{\DET}{\mbox{\rm DET}}
\newcommand{\hol}{\mbox{\rm hol}}
\newcommand{\End}{\mbox{\rm End}}
\newcommand{\spec}{\mbox{\rm spec}}
\renewcommand{\sf}{\mbox{\rm sf}}
\newcommand{\ind}{\mbox{\rm ind}}
\newcommand{\rk}{\mbox{\rm rk}}
\newcommand{\re}{\mbox{\rm Re}}
\newcommand{\Res}{\mbox{\rm Res}}
\newcommand{\HH}{{\rm H}}
\newcommand{\im}{\mbox{\rm Im}}
\newcommand{\inte}{\mbox{\rm int}}
\newcommand{\zn}{\stackrel{\circ}{\nabla}}
\newcommand{\bn}{\bar{\nabla}}
\newcommand{\zg}{\stackrel{\circ}{g}}

\newcommand{\Dirac}{\mathcal{D}}
\newcommand{\R}{\mathcal{R}}

\renewcommand{\theequation}{\arabic{section}.\arabic{equation}}
\newcommand{\sect}[1]{\section{#1} \setcounter{equation}{0}}

\newtheorem{theo}{Theorem}[section]

\newtheorem{lem}[theo]{Lemma}
\newtheorem{prop}[theo]{Proposition}
\newtheorem{coro}[theo]{Corollary}

\title{\sc A Note on Positive Energy Theorem for Spaces with Asymptotic SUSY Compactification}
\author{Xianzhe Dai}
\maketitle

\begin{abstract} We extend the positive mass theorem in \cite{d} to the Lorentzian
setting. This includes the original higher dimensional Positive
Energy Theorem whose spinor proof is given in \cite{Wi} and
\cite{PT} for dimension $4$ and in \cite{Z} for dimension $5$.
\end{abstract}

\section{Introduction and statement of the result}
In this note, we formulate and prove the Lorentzian version of the
positive mass theorem in \cite{d}. There we prove a positive mass
theorem for spaces which asymptotically approach the product of a flat Euclidean
space with a compact manifold which admits a nonzero parallel spinor (such as a
Calabi-Yau manifold or any special honolomy manifold except the
quaternionic K\"ahler). This is motivated by string theory,
especially the recent work \cite{HHM}. The application of the positve mass theorem of \cite{d} to the study of stability of
Ricci flat manifolds is discussed in \cite{dww}.

In general relativity, a spacetime is modeled by a Lorentzian
$4$-manifold $(N, \ g)$ together with an energy-momentum tensor
$T$ satisfying Einstein equation \be \label{ee} R_{\alpha\beta} -
\hf g_{\alpha \beta} R = 8\pi T_{\alpha \beta}. \ee The positive
energy theorem \cite{SY1}, \cite{Wi} says that an isolated
gravitational system with nonnegative local matter density must
have nonnegative total energy, measured at spatial infinity. More
precisely, one considers a complete oriented spacelike
hypersurface $M$ of $N$ satisfying the following two conditions:

a). {\em $M$ is asymptotically flat}, that is, there is a compact set
$K$ in $M$ such that $M-K$ is the disjoint union of a finite
number of subsets $M_1, \dots , M_k$ and each $M_l$ is
diffeomorphic to $({\Bbb R}^3 -B_R(0))$. Moreover, under this
diffeomorphism, the metric of $M_l$ is of the form \be \label{afm}
g_{ij}=\delta_{ij} + O(r^{-\tau}), \ \ \  \partial_k
g_{ij}=O(r^{-\tau -1}), \ \ \
\partial_k \partial_l g_{ij}=O(r^{-\tau -2}). \ee Furthermore, the
second fundamental form $h_{ij}$ of $M$ in $N$ satisfies \be
\label{affsff} h_{ij}= O(r^{-\tau -1}), \ \ \ \  \partial_k
h_{ij}=O(r^{-\tau -2}). \ee Here $\tau>0$ is the asymptotic order
and $r$ is the Euclidean distance to a base point.

b). {\em $M$ has nonnegative local mass density}: for each point $p \in
M$ and for each timelike vector $e_0$ at $p$, $T(e_0, \ e_0) \geq
0$ and $T(e_0, \ \cdot)$ is a nonspacelike co-vector. This implies
the dominant energy condition
\be \label{dec} T^{00} \geq |
T^{\alpha \beta} |, \ \ \ T^{00} \geq (-T_{0i}T^{0i})^{\hf}. \ee

The total energy (the ADM mass) and the total (linear) momentum of
$M$ can then be defined as follows \cite{ADM}, \cite{PT} (for
simplicity we suppress the dependence here on $l$ (the end $M_l$))
\ba \label{admm} E = \lim_{R \ra \infty} \frac{1}{4 \omega_n}
\int_{S_R} (\partial_i g_{ij} - \pt_j g_{ii} ) * dx_j , \nonumber \\
 P_k =\lim_{R \ra \infty} \frac{1}{4 \omega_n} \int_{S_R} 2(h_{jk} -
\delta_{jk} h_{ii} ) * dx_j \ea
Here $\omega_n$ denotes the volume
of the $n-1$ sphere and $S_R$ the Euclidean sphere with radius $R$
centered at the base point.

\begin{theo}[Schoen-Yau, Witten] With the assumptions as above and assuming that $M$ is spin, one has
\[ E - |P| \geq 0\]
on each end $M_l$. Moreover, if $E=0$ for some end $M_l$, then $M$
has only one end and $N$ is flat along $M$.
\end{theo}

Now, according to string theory \cite{CHSW}, our universe is really ten
dimensional, modelled on  $\mathbb R^{3,1} \times X$ where $X$ is a
Calabi-Yau 3-fold. This is the so called Calabi-Yau
compactification, which motivates the spaces we now consider.

Thus, we consider a Lorentzian manifold $N$ (with signature $(-,
+, \cdots, +)$) of $\dim N=n+1$, with a energy-momentum tensor
satisfying the Einstein equation. Then let $M$ be a complete
oriented spacelike hypersurface in $N$. Furthermore the Riemannian
manifold $(M^n, g)$ with $g$ induced from the Lorentzian metric
decomposes $M=M_0 \cup M_{\infty}$, where $M_0$ is compact as
before but now $M_{\infty} \simeq ({\Bbb R}^k - B_R(0)) \times X$
for some radius $R>0$ and $X$ a compact simply connected spin
manifold which admits a nonzero parallel spinor. Moreover the
metric on $M_{\infty}$ satisfies \be \label{afmsu} g=\zg + u, \ \
\ \zg=g_{{\mathbb R}^k} + g_X, \ \ \ u= O(r^{-\tau}), \ \ \ \zn\!
u=O(r^{-\tau -1}), \ \ \ \zn \zn \! u=O(r^{-\tau -2}), \ee and the
second fundamental form $h$ of $M$ in $N$ satisfies \be
\label{afsfsu} h=O(r^{-\tau-1}), \ \ \ \zn\! h=O(r^{-\tau -2}).
\ee Here $\zn$ is the Levi-Civita connection of $\zg$ (extended to
act on all tensor fields), $\tau
>0$ is the asymptotical order.

The total energy and total momentum for such a space can then be
defined by
\ba \label{admkkm}
E= \lim_{R \ra \infty} \frac{1}{4 \omega_k vol(X)} \int_{S_R \times X} (\partial_i g_{ij} - \pt_j g_{aa} ) * dx_j dvol(X),
\nonumber \\
P_k = \lim_{R \ra \infty} \frac{1}{4 \omega_k vol(X)} \int_{S_R
\times X} 2(h_{jk} - \delta_{jk} h_{ii} ) * dx_j dvol(X). \ea Here
the $*$ operator is the one on the Euclidean factor, the index $i,
j$ run over the Euclidean factor while the index $a$ runs over the
full index of the manifold.

Then we have

\begin{theo}
Assuming that $M$ is spin, one has
\[ E - |P| \geq 0\]
on each end $M_l$. Moreover, if $E=0$ for some end $M_l$, then $M$
has only one end. In this case, when $k=n$, $N$ is flat along $M$.
\end{theo}

In particular, this result includes the original higher
dimensional Positive Energy Theorem whose spinor proof is given in
\cite{Wi} and \cite{PT} for dimension $4$ and in \cite{Z} for
dimension $5$. 

{\em Acknowledgement:} This work is motivated and inspired by the
work of Gary Horowitz and his collaborators \cite{HHM}. The author
is indebted to Gary for sharing his ideas and for interesting
discussions. The author would also like to thank Xiao Zhang and Siye Wu for
useful discussion.

\section{The hypersurface Dirac operator}

We will adapt Witten's spinor method \cite{Wi}, as given in \cite{PT}, to our situation.
The crucial ingredient here is the hypersurface Dirac operator on
$M$, acting on the (restriction of the) spinor bundle of $N$. Let
$S$ be the spinor bundle of $N$ and still denote by the same
notation its restriction on (or rather, pullback to) $M$. Denote by $\nabla$
the connection on $S$ induced by the Lorentzian metric on $N$. The Lorentzian
metric on $N$ also induces a Riemannian metric on $M$, whose Levi-Civita
connection gives rise to another connection, $\bar{\nabla}$ on $S$.
The two, of course, differ by a term involving the second fundamental form.

There are two choices of metrics on $S$, which is another subtlety
here. Since part of the treatment in \cite{PT} is special to
dimension $4$, we will give a somewhat detailed account here.

Let $SO(n,1)$ denote the identity component of the groups of
orientation preserving isometries of the Minkowski space $\mathbb
R^{n,1}$. A choice of a unit timelike covector $e^0$ gives rise to
injective homomorphisms $\alpha$, $\hat{\alpha}$,  and a
commutative diagram \be
\label{injh} \begin{array}{cccc} \alpha: \ & SO(n) & \ra & SO(n,1) \\
& \uparrow & & \uparrow \\
\hat{\alpha}: \ & Spin(n) & \ra & Spin(n,1). \end{array} \ee

We now fix a choice of unit timelike normal covector $e^0$ of $M$
in $N$. Let $F(N)$ denote the $SO(n,1)$ frame bundle of $N$ and
$F(M)$ the $SO(n)$ frame bundle of $M$. Then
$i^*F(N)=F(M)\times_{\alpha} SO(n,1)$, where $i: \ M
\hookrightarrow N$ is the inclusion. If $N$ is spin, then we have
a principal $Spin(n,1)$ bundle $P_{Spin(n,1)}$ on $N$, whose
restriction on $M$ is then $i^*P_{Spin(n,1)}=P_{Spin(n)}
\times_{\hat{\alpha}} Spin(n,1)$, where $P_{Spin(n)}$ is the
principal $Spin(n)$ bundle of $M$. Thus, even if $N$ is not spin,
$i^*P_{Spin(n,1)}$ is still well-defined as long as $M$ is spin.

Similarly, when $N$ is spin, the spinor bundle $S$ on $N$ is the
associated bundle $P_{Spin(n,1)} \times_{\rho_{n,1}} \Delta$,
where $\Delta=\mathbb C^{2^{[\o{n+1}{2}]}}$ is the complex vector
space of spinors and \be \rho_{n,1}: \ Spin(n,1) \ra GL(\Delta)
\ee is the spin representation. Its restriction to $M$ is given by
$i^*P_{Spin(n,1)} \times_{\rho_{n,1}} \Delta=
P_{Spin(n)} \times_{\rho_{n}} \Delta$ with \be \rho_{n}: \
Spin(n) \stackrel{\hat{\alpha}}{\hookrightarrow} Spin(n,1)
\stackrel{\rho_{n,1}}{\longrightarrow} GL(\Delta) \ee Again, the restriction
is still well defined as long as $M$ is spin.

Let $e^0, e^i$ ($i=1, \cdots, n$ will be the range for the index $i$ in this section) be an orthonormal basis of the
Minkowski space $\mathbb R^{n,1}$ of dimension $n+1$ such that $|e^0|=-1$.

\begin{lem} There is a positive definite hermitian inner product $\lp \ , \ \rp$  on
$\Delta$ which is $Spin(n)$-invariant.  Moreover, $(s, s')= \lp
e^0 \cdot s, s' \rp$ defines a  hermitian inner product which is
also $Spin(n)$-invariant but not positive definite. In fact
\[ (v\cdot s, \ s')=(s, \ v\cdot s') \]
for all $v\in \mathbb R^{n,1}$.
\end{lem}

\Pf Detailed study via $\Gamma$ matrices \cite[p10-11]{CBDM} shows that
there is a positive definite hermitian inner product $\lp \ , \
\rp$ on $\Delta$ with respect to which $e^i$ is skew-hermitian
while $e^0$ is hermitian. It follows then that $\lp \ , \ \rp$ is
$Spin(n)$-invariant. We now show that $(s, s')=\lp e^0 \cdot s, s'
\rp $ defines a $Spin(n)$-invariant hermitian inner product. Since
$e^0$ is hermitian with respect to $\lp \ , \ \rp$, $(\ , \ )$ is
clearly hermitian. To show that $(\ , \ )$ is $Spin(n)$-invariant,
we take a unit vector $v$ in the Minkowski space: $v=a_0 e^0 + a_i
e^i$,  $a_0, a_i \in \mathbb R$ and $-a_0^2 + \sum_{i=1}^{n} a_i^2
=1$. Then
\ban (vs, vs') & =  & \lp e^0 vs, vs' \rp \\
& =  & a_0^2 \lp e^0 e^0s, e^0s' \rp + a_i a_0 \lp e^0 e^is, e^0s' \rp + a_0 a_i \lp e^0 e^0s, e^is' \rp +  a_ia_j \lp e^0
e^is, e^js' \rp\\
 & = & a_0^2 \lp s, e^0 s' \rp - a_ia_j \lp e^je^0 e^is, s' \rp \\
& = & a_0^2 \lp e^0 s, s' \rp + a_ia_j \lp e^0 e^je^is, s' \rp \\
& = & a_0^2 \lp e^0 s, s' \rp - a_i^2 \lp e^0 s, s' \rp \\
& = & -(s, s') \ean Consequently, $(\ , \ )$ is
$Spin(n)$-invariant.  The above computation also implies that
$v\cdot$ acts as hermitian operator on $\Delta$ with respect to
$(\ , \ )$. \qed

Thus the spinor bundle $S$ restricted to $M$ inherits an hermitian
metric $(\ , \ )$ and a positive definite metric $\lp \ , \ \rp$.
They are related by the equation \be (s, s')=\lp e^0 \cdot s, s'
\rp . \ee

Now the hypersurface Dirac operator is
defined by the composition

\be \label{hd} \Dirac: \ \Gamma(M, S)
\stackrel{\nabla}{\longrightarrow} \Gamma(M, T^*M \otimes S)
\stackrel{c}{\longrightarrow} \Gamma(M, S), \ee where $c$ denotes
the Clifford multiplication. In terms of a local orthonormal basis
$e_1, e_2, \cdots, e_n$ of $TM$,
\[ \Dirac \psi = e^i \cdot \nabla_{e_i} \psi,\]
where $e^i$ denotes the dual basis.

The two most important properties of hypersurface Dirac operator
are the self-adjointness with respect to the metric $\lp \ , \
\rp$ and the Bochner-Lichnerowicz-Weitzenbock formula \cite{Wi},
\cite{PT}.

\begin{lem} \label{fsa} Define a $n-1$ form on $M$ by $\omega=\lp \phi, e^i \cdot \psi \rp \inte(e_i) \, dvol$, where $dvol$
is the volume form of the Riemannian metric $g$. We have
\[ [ \lp \phi, \Dirac \psi \rp - \lp \Dirac \phi,  \psi \rp ]
dvol=d\omega. \] Thus $\Dirac$ is formally self adjoint with
respect to the $L^2$ metric defined by $\lp \ , \ \rp$ (and
$dvol$).
\end{lem}

\Pf Since $\omega$ is independent of the choice of the orthonormal basis, we do our computation locally using a preferred
basis. For any given point $p\in M$, choose a local orthonormal frame
$e_i$ of $TM$ near $p$ such that $\bar{\nabla}e_i=0$ at $p$.
Extend $e_0, e_i$ to a neighborhood of $p$ in $N$ by parallel
translating along $e_0$ direction. Then, at $p$,
$\nabla_{e_i}e^j=- h_{ij} e^0$ and $\nabla_{e_i}e^0=-h_{ij} e^j$.
Therefore (again at $p$), \ban d\omega & = & \nabla_{e_i} \lp
\phi, e^i \cdot \psi \rp \, dvol \\
& = & [((\nabla_{e_i}e^0)\cdot \phi, e^i \cdot \psi) + (e^0\cdot \nabla_{e_i} \phi, e^i \cdot \psi)
+ (e^0 \cdot \phi, (\nabla_{e_i}e^i) \cdot \psi) + (e^0 \cdot \phi, e^i \cdot \nabla_{e_i} \psi) ] dvol \\
& = & [ - h_{ij} (e^j \cdot \phi, e^i \cdot \psi) + (e^i\cdot
e^0\cdot \nabla_{e_i} \phi,  \psi ) - h_{ii} (e^0
\cdot \phi, e^0 \cdot \psi) + \lp \phi, \Dirac \psi \rp ] dvol \\
& = & [- h_{ij} (e^i \cdot e^j \cdot \phi,  \psi) - \lp \Dirac
\phi, \psi \rp - h_{ii} (e^0
\cdot \phi, e^0 \cdot \psi)   + \lp \phi, \Dirac \psi \rp ] dvol \\
& = & [- \lp \Dirac \phi, \psi \rp + \lp \phi, \Dirac \psi \rp ]
dvol \ean \qed

Now the Bochner-Lichnerowicz-Weitzenbock formula.

\begin{lem} One has
\ba \label{wlf} \Dirac^2=\nabla^* \nabla + \R,\hspace{1in} \\
\R=\o{1}{4}(R + 2R_{00} + 2R_{0i}e^0\cdot e^i \cdot ) \in End(S).
\nonumber \ea Here the adjoint $\nabla^*$ is with respect to the
metric $\lp \ , \ \rp$.
\end{lem}

\Pf We again do the computation in the frame as in the proof of
Lemma \ref{fsa}. Then \ban \Dirac^2 & = & e^i \cdot e^j \cdot
\nabla_{e_i} \nabla_{e_j} + e^i \cdot \nabla_{e_i}e^j \cdot
\nabla_{e_j} \\
& = & - \nabla_{e_i} \nabla_{e_i} + \o{1}{4} (R + 2R_{00} +
2R_{0i}e^0\cdot e^i \cdot ) - h_{ij} e^i \cdot e^0\cdot
\nabla_{e_j}. \ean Now \ban d [\lp \phi, \psi \rp \inte(e_i) \,
dvol ] & = & e_i \lp \phi, \psi \rp \, dvol \\
& = & (\nabla_{e_i}e^0 \cdot \phi, \psi) + \lp \nabla_{e_i} \phi,
\psi \rp + \lp  \phi, \nabla_{e_i} \psi \rp \\
& = & - h_{ij} (e^j \cdot \phi, \psi)
 + \lp \nabla_{e_i} \phi,
\psi \rp + \lp  \phi, \nabla_{e_i} \psi \rp \\
& = & - h_{ij} \lp e^0 \cdot e^j \cdot  \phi, \psi \rp + \lp
\nabla_{e_i} \phi, \psi \rp + \lp  \phi, \nabla_{e_i} \psi \rp
\ean This shows that $\nabla^*_{e_i}= - \nabla_{e_i} -h_{ij}e^j
\cdot e^0 \cdot$. The desired formula follows. \qed

\section{Proof of the Theorem}

By the Einstein equation,
\[ \R=4\pi (T_{00} + T_{0i}e^0\cdot e^i \cdot).\]
It follows then from the dominant energy condition (\ref{dec}) that

\be \label{decc} \R \geq 0. \ee

Now, for $\phi \in \Gamma(M, S)$ and a compact domain $\Omega
\subset M$ with smooth boundary, the Bochner-Lichnerowicz-Weitzenbock formula yields

\ba \label{iwf}
 \int_{\Omega} [  |\nabla \phi|^2 + \lp \phi, \R\phi \rp - |\Dirac \phi|^2
]\, dvol(g) & = & \int_{\pt \Omega} \sum \lp (\nabla_{e_a} +
e_a \cdot \Dirac)\phi, \ \phi \rp \, \inte(e_a) \, dvol(g) \ \ \ \ \ \\
& = & \int_{\pt \Omega} \sum \lp (\nabla_{\nu} + \nu \cdot \Dirac
)\phi, \ \phi \rp \, dvol(g|_{\pt \Omega}), \ \ \ \ \ \ea where
$e_a$ is an orthonormal basis of $g$ and $\nu$ is the unit outer
normal of $\pt \Omega$. Also, here $\inte(e_a)$ is the interior
multiplication by $e_a$.

Now let the manifold $M= M_0 \cup M_{\infty}$ with $M_0$ compact
and $M_{\infty} \simeq ({\Bbb R}^k - B_R(0)) \times X$, and $(X,
g_X)$ a compact Riemannian manifold with nonzero parallel spinors.
Moreover, the metric $g$ on $M$ satisfies (\ref{afmsu}). Let
$e^0_a$ be the orthonormal basis of $\zg$ which consists of
$\o{\pt}{\pt x_i}$ followed by an orthonormal basis $f_{\alpha}$
of $g_X$. Orthonormalizing $e^0_a$ with respect to $g$ gives rise
an orthonormal basis $e_a$ of $g$. Moreover, \be \label{onbasy}
e_a=e^0_a - \hf u_{ab} e^0_b + O(r^{-2\tau}). \ee This gives rise
to a gauge transformation
\[ A: \ SO(\zg) \ni e^0_a \ra e_a \in SO(g) \]
which identifies the corresponding spin groups and spinor bundles.

We now pick a unit norm parallel spinor $\psi_0$ of $({\Bbb R}^k,
g_{{\mathbb R}^k})$ and a unit norm parallel spinor $\psi_1$ of
$(X, g_X)$. Then $\phi_0=A(\psi_0 \otimes \psi_1)$ defines a
spinor of $M_{\infty}$.  We extend $\phi_0$ smoothly inside. Then
$\nabla^0 \phi_0 =0$ outside the compact set.

\begin{lem} \label{comt}
If a spinor $\phi$ is asymptotic to $\phi_0$: $\phi=\phi_0 + O(r^{-\tau})$, then we have
\[ \lim_{R \ra \infty} \Re \int_{S_R \times X} \sum \lp (\nabla_{e_a} + e_a \cdot D)
\phi, \ \phi \rp \, \inte(e_a) \, dvol(g) = \omega_k vol(X) \lp
\phi_0, \ E \phi_0 + P_k dx^0 \cdot dx^k \cdot \phi_0 \rp,
\]
where $\Re$ means taking the real part.
\end{lem}

\Pf Recall that $\bn$ denote the connection on $S$ induced from the
Levi-Civita connection on $M$. We have

\be \label{2con} \nabla_{e_a} \psi = \bn_{e_a} \psi -\hf h_{ab}e^0
\cdot e^b \cdot \psi. \ee By the Clifford relation,
\[ \lp (\nabla_{e_a} + e_a \cdot D)
\phi, \ \phi \rp=-\hf \lp [e^a\cdot, e^b\cdot]\nabla_{e_b} \phi, \
\phi \rp. \] Hence \ban \int_{S_R \times X} \sum \lp (\nabla_{e_a}
+ e_a \cdot D) \phi, \ \phi \rp \, \inte(e_a) \, dvol(g)  = \hspace{3in} \\
 -\hf \int_{S_R \times X}\lp [e^a\cdot, e^b\cdot]\bn_{e_b} \phi, \
\phi \rp\, \inte(e_a) \, dvol(g) +  \o{1}{4}\int_{S_R \times X}\lp
[e^a\cdot, e^b\cdot]h_{bc} e^0 \cdot e^c \cdot \phi, \ \phi \rp\,
\inte(e_a) \, dvol(g). \ean Using (\ref{onbasy}) and the asymptotic conditions (\ref{afsfsu}), the second term in the right
hand side can be easily seen to give us
\[ \lim_{R \ra \infty} \o{1}{4} \int_{S_R \times X} \lp 2(h_{ac}-\delta_{ac}h_{bb}) e^0 \cdot e^c \cdot \phi, \ \phi \rp\,
\inte(e_a)
\, dvol(g) = \omega_k vol(X) \lp \phi_0, \  P_k dx^0 \cdot dx^k
\cdot \phi_0 \rp. \]

The first term is computed in \cite{d} to limit to
\[ \omega_k vol(X) \lp \phi_0, \ E \phi_0 \rp.\]
\qed

The following lemma is standard \cite{PT}, \cite{Wi}.

\begin{lem} If
\[ \lp \phi_0, \ E \phi_0 + P_k dx^0 \cdot dx^k \cdot \phi_0 \rp
\geq 0 \] for all constant spinors $\phi_0$, then
\[ E - |P| \geq 0.\]
\end{lem}

As usual, the trick to get the positivity now is to find a
harmonic spinor $\phi$ asymptotic to $\phi_0$. Then the left hand
side of (\ref{iwf}) will be nonnegative since $\R \geq 0$. Passing
to the right hand side will give us the desired result.

\begin{lem} \label{hspinor}
There exists a harmonic spinor $\phi$ on $(M, \ g)$ which is
asmptotic to the parallel spinor $\phi_0$ at infinity:
\[ \Dirac \phi =0, \ \ \  \phi=\phi_0 + O(r^{-\tau}).  \]
\end{lem}

\Pf The proof is essentially the same as in \cite{d}. We use the
Fredholm property of $\Dirac$ on a weighted Sobolev space
and $\R \geq 0$ to show that it is an isomorphism. The harmonic
spinor $\phi$ can then be obtained by setting $\phi=\phi_0 + \xi$
and solving $\xi \in O(r^{-\tau})$ from the equation $\Dirac \xi =
-\Dirac \phi_0$. \qed

The rest of the Theorem follows as in \cite{PT}.

\end{document}